\documentstyle{article}
\input{tcilatex}
\begin{document}
\title{Approximate Symmetry Analysis of \\
a Class of Perturbed Nonlinear Reaction-Diffusion Equations}
\author{Mehdi Nadjafikah\thanks{%
Department of Mathematics, Islamic Azad University, Karaj Branch, P. O. Box
31485--313, Karaj, I. R. Iran. email: m\_nadjafikhah@iust. ac. ir} \and %
Abolhassan Mahdavi\thanks{%
e-mail: ad.mahdavi@kiau.ac.ir}}
\maketitle
\begin{abstract}
In this paper, the problem of approximate symmetries of a class of nonlinear reaction-diffusion equations called Kolmogorov-Petrovsky-Piskounov (KPP)
equation is comprehensively analyzed. In order to compute the approximate symmetries, we have applied the method which was proposed by Fushchich and
Shtelen \cite{[3]} and fundamentally based on the expansion of the dependent variables in a perturbation series. Particularly, an optimal system of one
dimensional subalgebras is constructed and some invariant solutions corresponding to the resulted symmetries are obtained. \\[2mm]
\end{abstract}
%%%%%%%%%%%%%%%%%%%%%%%%%%%%%%%%%%%%%
\section{Introduction}
Nonlinear problems arise widely in various fields of science and engineering mainly due to the fact that most physical systems are inherently nonlinear in nature. But for nonlinear partial differential equations (PDEs), analytical solutions are rare and difficult to obtain. Hence, the investigation of the exact solutions of nonlinear PDEs plays a fundamental role in the analysis of nonlinear physical phenomena. One of the most famous and established procedures for obtaining exact solutions of differential equations is the classical symmetries method, also called group analysis. This method was originated in 1881 from the pioneering work of Sophus Lie  \cite{[Lie]}. The investigation of symmetries has been manifested as one of the most significant and fundamental methods in almost every branch of science such as in mathematics and physics. Nowadays, the application of Lie group theory for the construction of solutions of nonlinear PDEs can be regarded as one of the most active fields of research in the theory of nonlinear PDEs and many good books have been dedicated to this subject (such as \cite{[6],[5],[olver]}). For some nonlinear problems, however symmetries are not rich to determine useful solutions. Hence, this fact was the motivation for the creation of several generalizations of the classical Lie group method. Consequently, several alternative reduction methods have been introduced, going beyond Lie's classical procedure and providing further solutions. One of the techniques widely applied in analyzing nonlinear problems is the perturbation analysis. Perturbation theory comprises mathematical methods that are applied to obtain an approximate solution to a problem which can not be solved exactly. Indeed, this procedure is performed by expanding the dependent variables asymptotically in terms of a small parameter. In order to combine the power of the Lie group theory and perturbation analysis, two different approximate symmetry theories have been developed recently. The first method is due to Baikov, Gazizov and Ibragimov  \cite{[baikov],[1]}. Successively another method for obtaining approximate symmetries was introduced by Fushchich and Shtelen \cite{[3]}.

In the method proposed by Baikov, Gazizov and Ibragimov, the Lie operator is expanded in a perturbation series other than perturbation for dependent variables as in the usual case. In other words, assume that the perturbed differential equation be in the form: $F(z)=F_{0}(z)+\varepsilon F_{1}(z)$, where $z=(x,u,u_{(1)},\cdots ,u_{(n)})$, $F_{0}$ is the unperturbed equation, $F_{1}(z)$ is the perturbed term and $X=X^{0}+\varepsilon X^{1}$ is the corresponding infinitesimal generator. The exact symmetry of the unperturbed equation $F_{0}(z)$ is denoted by $X^{0}$ and can be obtained as $X^{0}F_{0}(z)\left\vert _{F_{0}(z)=0}\right. =0$. Then, by applying the auxiliary function $H=\frac{1}{\varepsilon }X^{0}\left( F_{0}(z)+\varepsilon F_{1}(z)\right)\big\vert _{F_{0}+\varepsilon F_{1}=0}$,  
vector field $X_{1}$ will be deduced from the following relation: 
\begin{equation}
X^{1}F_{0}(z)\big\vert_{F_{0}=0}+H=0.
\end{equation}

Finally, after obtaining the approximate symmetries, the corresponding
approximate solutions will be obtained via the classical Lie symmetry method 
\cite{[8]}. \newline
In the second method due to Fushchich and Shtelen, first of all the
dependent variables are expanded in a perturbation series. In the next step,
terms are then separated at each order of approximation and as a consequence
a system of equations to be solved in a hierarchy is determined. Finally,
the approximate symmetries of the original equation is defined to be the
exact symmetries of the system of equations resulted from perturbations \cite%
{[4], [3], [7]}. Pakdemirli et al. in a recent paper \cite{[compare]} have
compared these above two methods. According to their comparison, the
expansion of the approximate operator applied in the first method, does not
reflect well an approximation in the perturbation sense; While the second
method is consistent with the perturbation theory and results correct terms
for the approximate solutions. Consequently, the second method is superior
to the first one according to the comparison in \cite{[compare]}. \newline
$~~$Nonlinear reaction-diffusion equations can be regarded as mathematical
models which explain the change of the concentration of one or more
substances distributed in space. Indeed, this variation occurs under the
influence of two main processes including chemical reactions in which the
substances are locally transformed into each other, and diffusion which
makes the substances to spread out over a surface in space. From the
mathematical point of view, reaction-diffusion systems generally take the
form of semi-linear parabolic PDEs. It is worth mentioning that the
solutions of reaction-diffusion equations represent a wide range of
behaviors, such as formation of wave-like phenomena and traveling waves as
well as other self-organized patterns. \newline
In this paper, we will apply the method proposed by Fushchich and Shtelen 
\cite{[3]} in order to present a comprehensive analysis of the approximate
symmetries of a significant class of nonlinear reaction-diffusion equations
called Kolmogrov-Petrovsky-Piskounov (KPP) equation \cite{[kpp]}. This
equation can be regarded as the most simple reaction-diffusion equation
concerning the concentration $u$ of a single substance in one spatial
dimension and is generally defined as follows: 
\begin{equation}
u_{t}-u_{xx}=R(u).  \label{kpp}
\end{equation}%
By inserting different values to the reaction term $R(u)$ of equation (\ref%
{kpp}), the following significant equations are deduced:
\begin{itemize}
\item[1. ] If the reaction term $R(u)$ vanishes, then the resulted equation
displays a pure diffusion process and defined by: 
\begin{equation}  \label{fick}
u_t=u_{xx},
\end{equation}
Note that the above equation is called Fick's second law \cite{[kpp]}.
\item[2. ] By inserting $R(u)=au(1-u),\ a\geq 0,$ the Fisher equation (or
logistic equation) is resulted as follows: 
\begin{equation}  \label{fisher}
u_t=u_{xx}+au(1-u),
\end{equation}
This equation can be regarded as the archetypical deterministic model for
the spread of a useful gene in a population of diploid individuals living in
a one dimensional habitat \cite{[fisher1],[fisher2]}.
\item[3. ] By inserting $R(u)=u^{2}(1-u)$, the Zeldovich equation will be
deduced as follows: 
\begin{equation}  \label{zeld}
u_t=u_{xx}+u^2(1-u),
\end{equation}
This equation appears in combustion theory. The unknown $u$ displays
temperature, while the last term on the right-hand side is concerned to the
generation of heat by combustion \cite{[zeld],[zeld2]}.
\item[4. ] By inserting $R(u)=u(1-u^{2})$ the Newell-Whitehead-Segel (NWS)
equation (or amplitude equation) is resulted as follows: 
\begin{equation}  \label{new}
u_t=u_{xx}+u(1-u^2).
\end{equation}
This equation arises in the analysis of thermal convection of a fluid heated
from below after carrying out a suitable normalization \cite{[new]}.
\end{itemize}
$~~$ This paper is organized as follows: Section 2 is devoted to the
thorough investigation of the approximate symmetries and approximate
solutions of the KPP equation. For this purpose, we will concentrate on the
four special and significant forms of the KPP equation described above i.e
Fick's second law, Fisher's equation, Zeldovich equation and
Newell-Whitehead-Segel (NWS) equation. In section 3, an optimal system of
subalgebras is constructed and the corresponding symmetry transformations
are obtained. Some concluding remarks are mentioned at the end of the paper. 
%%%%%%%%%%%%%%%%%%%%%%%%%%%%%%%%%%%%%%%%%%%%%%%%%%%%%%%%%%
%%%%%%%%%%%%%%%%%%%%%%%%%%%%%%%%%%%%%%%%%%%%%%%%%%%%%%%%%%%%%%%%%%%%%
\section{Approximate symmetries of the KPP equation}
In this section, first of all the problem of exact and approximate
symmetries of the Fick's second law (\ref{fick}) with a small parameter is
investigated. Then the approximate symmetries and the exact and approximate
invariant solutions corresponding to the perturbed Fisher's equation,
Zeldovich equation and Newell-Whiehead-Segel (NWS) equation will be
determined.
%%%%%%%%%%%%%%%%%%%%%%%%%%%%%%%%%%%%%%%%%%%%%%%%%%%%%%%%%%%%%
\subsection{Exact symmetries of the perturbed Fick's second law}
The perturbed Fick's second law is defined as follows: 
\begin{equation}
u_{t}=\varepsilon u_{xx}.  \label{eq:4}
\end{equation}%
where $\varepsilon $ is a small parameter. Let $X$ be the infinitesimal
symmetry generator corresponding to the equation (\ref{eq:4}) which is
defined as follows: 
\begin{equation}\label{eq:5}
X=\xi (x,t,u)\partial_x+\tau (x,t,u)\,\partial_t+\varphi (x,t,u)\partial_u.  
\end{equation}%
Now by acting the second prolongation of the symmetry operator (\ref{eq:5})
on equation (\ref{eq:4}), an overdetermined system of equations for $\xi
,\tau $ and $\varphi $ will be obtained. By solving this resulted
determining equations, it is inferred that: 
\begin{eqnarray}\label{eq:6}
\xi &=&(c_{1}tx+c_{2}x)-2\varepsilon c_{4}t+c_{6},  \nonumber  \\
\tau &=&c_{1}t^{2}+2c_{2}t+c_{3}, \\
\qquad \varphi &=&(c_{4}x+c_{5}-c_{1}t/2-c_{1}x^{2}/4\varepsilon)u+F(x,t).  \nonumber
\end{eqnarray}%
where $F(x,t)$ is an arbitrary function satisfying the perturbed Fick's second law equation (\ref{eq:5}), and $c_{i},\ i=1,\cdots ,6$ are arbitrary constants. Hence, this equation admits a six-dimensional Lie algebra with the following generators: 
\begin{eqnarray}
\begin{array}{l}
\displaystyle X_{1}=\partial_x, \\[2mm] 
\displaystyle X_{2}=\,\partial_t, \\[2mm] 
\displaystyle X_{3}=x\,\partial_x+2t\,\partial_t,
\end{array} \qquad 
\begin{array}{l}
\displaystyle X_{4}=-2\varepsilon t\,\partial_x+xu\,\partial_u, \\[2mm] 
\displaystyle X_{5}=u\,\partial_u, \\[2mm] 
\displaystyle X_{6}=4xt\,\partial_x+4t^2\partial_t-(2t +x^2/\varepsilon)u\,\partial_u.%
\end{array}
\label{eq:7}
\end{eqnarray}%
plus the following infinite dimensional subalgebra which is spanned by $X_F=F(x,t)\,\partial_u$, where $F$ satisfies (\ref{eq:5}). 
%
%%%%%%%%%%%%%%%%%%%%%%%%%%%%%%%%%%%%%%%%%%%%%%%%%%%%%%%
%%%%%%%%%%%%%%%%%%%%%%%%%%%%%%%%%%%%%%%%%%%%%%%%%%%%%%%
%
\subsection{Exact invariant solutions}
In this part, we compute some exact invariant solutions corresponding to the resulted infinitesimal generators.
\paragraph{Case 1.}
Consider the symmetry operator $X=cX_{1}+X_{2}$, where $c$ is a constant.

Now taking into account \cite{[6], [5],[olver]}, by applying the Lie symmetry reduction technique the corresponding exact and approximate invariant solutions will be obtained as follows. The characteristic equation associated to the symmetry generator $X$ is given by  $dx/c=dt/1=du/0$. By solving above equation, the following Lie invariants are resulted: $x-ct=y$, $u=v(y)$. By substituting these invariants into equation (\ref{eq:4}) we obtain: 
$\varepsilon v^{\prime \prime }(y)+cv^{\prime }(y)=0$. Consequently, by solving the above resulted ODE, the following solution is
deduced for equation (\ref{eq:4}): $u(x,t)=c_{1}+c_{2}\exp(-c(x-ct)/\varepsilon)$.
\paragraph{Case 2.}
For the symmetry generator $X_{3}$, the corresponding characteristic equation is: $dx/x=dt/2t=du/0$. Thus, these Lie invariants are determined: 
$u=v(y)$, $y=x^2/t$. By substituting above invariants into (\ref{eq:4}) the following ODE is inferred: 
$4\varepsilon yv^{\prime \prime }(y)+v^{\prime }(y)(2\varepsilon +y)=0$. Hence, another solution is deduced for equation (\ref{eq:4}):  
$u=v(y)=c_{1}+c_{2}\,{\rm erf}(|x|/2\sqrt{\varepsilon t})$, where $c_{1}$ and $c_{2}$ are arbitrary constants, and erf is the error
function given by:  ${\rm erf}(x)=(2/\sqrt{\pi})\,\int_{0}^{x}e^{-t^{2}}dt$.
%
%
%%%%%%%%%%%%%%%%%%%%%%%%%%%%%%%%%%%%%%%%%%%%%%%%%%%%%%%%%%%%%%%%%%%%%%%
%%%%%%%%%%%%%%%%%%%%%%%%%%%%%%%%%%%%%%%%%%%%%%%%%%%%%%%%%%%%%%%%%%%%%%%%%%%55
\subsection{Perturbed Fisher's equation}
In this section, a thorough investigation of the symmetries of the perturbed Fisher's equation is proposed: 
\begin{equation}
u_{t}=\varepsilon u_{xx}+au(1-u)  \label{pfisher}
\end{equation}%
For this purpose, firstly the exact symmetries of the perturbed Fisher's equation (\ref{pfisher}) will be calculated. Then, the approximate symmetries of this equation will be analyzed.

Now by acting the second prolongation of the symmetry generator (\ref{eq:5}) on the perturbed Fisher's equation and solving the resulted determining equations, it is deduced that $\xi =c_{2}$, $\tau =c_{1}$, and $\varphi =0$. where $c_{1}$ and$\ c_{2}$ are arbitrary constants. Hence, the following exact trivial symmetries are obtained: $X_{1}=\partial_x$, $X_{2}=\partial_t$. For the infinitesimal symmetry generator $X=c\partial_x+\partial_t$, the corresponding characteristic equation is given by $dx/c=dt/1=du/0$.

Therefore, the Lie invariants are resulted as $x-ct=y$ and $u=v(y)$. After substituting these invariants into the perturbed Fisher's equation, the following reduced ordinary differential equation is obtained: 
\begin{equation}
\varepsilon v^{\prime \prime }(y)+cv^{\prime }(y)+av(y)(1-v(y))=0.
\label{eq:8}
\end{equation}%
But it is worth noting that finding an exact solution for the differential equation (\ref{eq:8}) is difficult. For the particular case $c=\pm 5/\sqrt{6}$, Ablowitz and Zepptella \cite{[9]} used Painleve's singularity structure analysis in order to obtain the first corresponding explicit analytical solution which is given by: 
\begin{equation}
v(y)=u(x,t)=\left[ 1+\frac{\varepsilon }{\sqrt{6}}\exp\Big(\sqrt{6}x-\frac{5}{6}t\Big)\right] ^{-2}.
\end{equation}
%%%%%%%%%%%%%%%%%%%%%%%%%%%%%%%%%%%%%%%%%%%%%%%%%%%%%%%%%%%%%%%%%%%%%%%%%%%%%%%%
%%%%%%%%%%%%%%%%%%%%%%%%%%%%%%%%%%%%%%%%%%%%%%%%%%%%%%%%%%%%%%%%%%%%%%%%%%%%%%%%%%%%%%%%%%%%%
\subsubsection{Approximate symmetries of the perturbed Fisher's equation}
In this section, we apply the method proposed in \cite{[3]} in order to
analyze the problem of approximate symmetries of the Fisher's equation with
an accuracy of order one. First, we expand the dependent variable in
perturbation series, and then we separate terms of each order of
approximation, so that a system of equations will be formed. The derived
system is assumed to be coupled and its exact symmetry will be considered as
the approximate symmetry of the original equation.

We expand the dependant variable up to order one as follows: 
\begin{equation}  \label{eq:9}
u=v+\varepsilon w, \qquad 0<\varepsilon \leq 1.
\end{equation}
Where $v$ and $w$ are smooth functions of $x$ and $t$. After substitution of (\ref{eq:9}) into the perturbed Fisher's equation (\ref{pfisher}) and
equating to zero the coefficients of $o(\varepsilon^{0})$ and $o(\varepsilon^{1})$, the following system of partial differential equations is resulted: 
\begin{eqnarray}  \label{eq:10}
O(\varepsilon^{0}):\,v_{t}-av(1-v)w=0,\quad O(\varepsilon ):\,w_{t}-v_{xx}-aw(1-2v)=0.
\end{eqnarray}
\paragraph{Definition:}
The approximate symmetry of the Fisher's equation with a small parameter is
called the exact symmetry of the system of differential equations (\ref%
{eq:10}).

Now, consider the following symmetry transformation group acting on the PDE system (\ref{eq:10}): 
\begin{eqnarray}
\widetilde{x}=x+a\xi _{1}(t,x,v,w)+o(a^{2}),\quad &&\widetilde{t}=t+a\xi_{2}(t,x,v,w)+o(a^{2}),  \nonumber \\
\widetilde{v}=v+a\varphi _{1}(t,x,v,w)+o(a^{2}),\quad &&\widetilde{w}=w+a\varphi _{2}(t,x,v,w)+o(a^{2}),
\end{eqnarray}%
where $a$ is the group parameter and $\xi _{1},\xi _{2}$ and $\varphi _{1},$  $\varphi _{2}$ are the infinitesimals of the transformations for the independent and dependent variables, respectively. The associated vector field is of the form: 
\begin{equation}
X=\xi _{1}(t,x,v,w)\partial_t+\xi _{2}(t,x,v,w)\,\partial_x+\varphi _{1}(t,x,v,w)\,\partial_v+\varphi _{2}(t,x,v,w)\partial_w.  \label{eq:11}
\end{equation}%
The invariance of the system (\ref{eq:10}) under the infinitesimal symmetry transformation group (\ref{eq:11}) leads to the following invariance condition: 
$pr^{(2)}X\left[ \Delta \right] =0$, and $\Delta =0$. Hence, the following set of determining equations is inferred: 
\begin{equation}
\partial _{w}\xi _{2}=0,\quad av^{2}\partial _{w}\xi _{1}+\partial _{w}\varphi _{1}-av\partial _{w}\xi _{1}=0,\quad \cdots ,\quad 2\partial_{vx}\xi _{2}-\partial _{vv}\varphi _{1}=0.
\end{equation}%
By solving this system of PDEs, it is deduced to:  $\xi _{2}=C_{1}x+C_{3}$, $\varphi _{1}=0$, $\xi _{1}=C_{2}$, $\varphi _{2}=-2C_{1}w$, where $C_{1}$, $C_{2}$ and $C_{3}$ are arbitrary constants. Thus, the Lie algebra of the resulted infinitesimal symmetries of the PDE system (\ref{eq:10}) is spanned by these three vector fields: 
\begin{equation}
X_{1}=\partial_t,\qquad X_{2}=\partial_x,\qquad X_{3}=x\partial_x-2w\partial_w.
\end{equation}
%%%%%%%%%%%%%%%%%%%%%%%%%%%%%%%%%%%%%%%%%%%%%%%%%%%%%%%%%%%%%%%%%%%%%%%%%%%%%%%%%%%%%%%%%
\subsubsection{Approximate invariant solutions}
In this section, the approximate solutions will be obtained from the
approximate symmetries which were resulted in the pervious section.
\paragraph{Case 1. $X=x\partial_x-2w\,\partial_w$.}
By applying the classical Lie symmetry group method, the corresponding characteristic equation is  $dx/x=dt/0=dv/0=dw/(-2w)$. So that the resulted invariants are: $t=T$, $v=f(T)$ and $w=g(T)/x^2$. After substituting these invariants into the first equation of the PDE system (\ref{eq:10}), we have: 
\begin{equation}
f^{\prime }(T)-af(T)(1-f(T))=0.  \label{eq:12}
\end{equation}
Consequently, the following solution is obtained: 
\begin{equation}
f(T)=v=1/(1+c_{1}e^{-at}).
\end{equation}
After substituting $v$ in the second equation of the PDE system (\ref{eq:10}), this ODE is resulted $g^{\prime }(T)+ag(T)\left[2/(1+c_{1}e^{-at})-1\right] =0$. 
Therefore, we have $g(T)=c_{2}e^{-at}/\left( 1+c_{1}e^{-at}\right)^2$. Finally, taking into account (\ref{eq:9}), the following approximate solution is inferred: 
\begin{equation}
u(x,t)=v+\varepsilon w=1/(1+c_{1}e^{-at})+\varepsilon\,c_{2}e^{-at}/(x^{2}\left( 1+c_{1}e^{-at}\right) ^{2}),
\end{equation}
where $c_{1}$ and $c_{2}$ are arbitrary constants. 
%%%%%%%%%%%%%%%%%%%%%%%%%%%%%%%%%%%%%%%%%%%%%%%%%%%%%%%%%%%%
%%%%%%%%%%%%%%%%%%%%%%%%%%%%%%%%%%%%%%%%%%%%%%%%%%%%%%%%%%%%%%%%%%%%%%%%
\paragraph{Case 2.}
Now consider $X=X_{1}+cX_{2}$ where $c$ is an arbitrary constant. The corresponding characteristic equation is defined by $dx/c=dt/1=dv/0=dw/0$. So, the associated Lie invariants are $x-ct=y$, $v=f(y)$, and $w=g(y)$. By substituting the resulted invariants into the first equation of the PDE system (\ref{eq:10}), the reduced equation is determined as $cf^{\prime }(y)+af(y)(1-f(y))=0$. Therefore, we have $v(x,t)=1/(c_{1}e^{a(x-ct)/c}+1)$. Now by substituting $v(x,t)$ into the second equation of the PDE system (\ref{eq:10}), it is inferred that: 
\begin{equation}
cg(y)+\frac{c_{1}a^2\,e^{ay/c}(-1+c_{1}e^{ay/c})}{c_2\left(1+c_{1}e^{ay/c}\right) ^{3}}+g(y)\left( 1-\frac{2}{c_{1}e^{ay/c}+1}\right)=0.
\end{equation}%
By solving above equation, we have: 
\begin{equation}
g(y)=\frac{e^{ay/c}}{\left( c_{1}e^{ay/c}+1\right) ^{2}}\,\Big(c_{1}\frac{a^{2}}{c_{3}}y-\frac{2ac_{1}}{c_{2}}\ln \left( c_{1}e^{ay/c}+1\right) +c_{2}\Big).
\end{equation}
Finally, the following approximate solution is resulted: 
\begin{eqnarray}
u(x,t) &=&v+\varepsilon w =\frac{1}{c_{1}e^{a(x-ct)/c}+1}\,\Big\{1+ \\
&&\varepsilon e^{a(x-ct)/c}\,\left( \frac{c_{1}}{c^{3}}(x-ct)-\frac{2ac_{1}}{c^{2}}\ln \left( c_{1}e^{a(x-ct)/c}+1\right) +c_{2}\right) \Big\}.  \nonumber
\end{eqnarray}%
Consequently, the approximate solutions corresponding to all the resulted
operators were computed. 
%%%%%%%%%%%%%%%%%%%%%%%%%%%%%%%%%%%%%%%%%%%%%%%%%%%%%%%%%%%%%
%%%%%%%%%%%%%%%%%%%%%%%%%%%%%%%%%%%%%%%%%%%%%%%%%%%%%%%%%%%%%%%%%%%%%%%

\subsection{Perturbed Zeldovich equation}

In this section, we will investigate the exact and approximate symmetries of the Zeldovich equation with a small parameter: 
\begin{equation}  \label{eq:13}
u_{t}-\varepsilon u_{xx}=u^2(1-u).
\end{equation}
For this purpose, first of all we will compute the exact symmetries, then by applying the classical Lie symmetry method, the perturbed Zeldovich equation would be converted to an ODE.

By acting the symmetry operator (\ref{eq:5}) on the perturbed Zeldovich equation (\ref{eq:13}) and solving the resulted determining equations we have: $\xi =c_{1}$, $\tau =c_{2}$, $\varphi =0$, where $c_{1}$ and $c_{2}$ are arbitrary constants. Hence, the corresponding infinitesimal symmetries will be spanned by these two vector fields $X_{1}=\partial_t$, and $X_{2}=\partial_x$.  The characteristic equation corresponding to the symmetry operator $X=X_{1}+cX_{2}$ is given by $dx/c=dt/1=du/0$. Hence, the Lie invariants are obtained as $x-ct=y$, and $u=f(y)$. After substituting these invariants into equation (\ref{eq:13}), the  reduced equation is inferred as $\varepsilon f^{\prime \prime }(y)+cf^{\prime 2}(y)(1-f(y))=0$.
%%%%%%%%%%%%%%%%%%%%%%%%%%%%%%%%%%%%%%%%%%%%%%%%%%%%%%%%%%%%%%%%%%%%%%%
%%%%%%%%%%%%%%%%%%%%%%%%%%%%%%%%%%%%%%%%%%%%%%%%%%%%%%%%%%%%%%%%%%%%%%%%%%%%%%%%
\subsubsection{Approximate Symmetries of the Zeldovich equation}
In this section, we use the method proposed in \cite{[3]} in order to obtain
the approximate symmetries of the equation (\ref{eq:13}) with the accuracy $%
o(\varepsilon )$. By expanding the dependent variable of this equation in
perturbation series we have: 
\begin{equation}
u=v+\varepsilon w,\qquad 0\leq \varepsilon \leq 1.
\end{equation}%
Then by substituting the above relation into the perturbed equation (\ref{eq:13}) and separating terms of each order of approximation, the following equations with respect to $o(\varepsilon ^{0})$ and $o(\varepsilon ^{1})$ are deduced: 
\begin{eqnarray}
O(\varepsilon ^{0}):\,v_{t}-v^{2}(1-v)=0,  \label{eq:14} \quad O(\varepsilon ^{1}):\,w_{t}-v_{xx}-2vw(1-v)+v^{2}w=0. 
\end{eqnarray}%
It is worth mentioning that the resulted approximate symmetries of the
differential equation (\ref{eq:13}) correspond to the exact symmetries of
the PDE system (\ref{eq:14}).

Now, by acting the second prolongation of the infinitesimal symmetry operator (\ref{eq:11}) on the PDE system (\ref{eq:14}) and solving the resulted determining equations, we have $\xi _{1}=c_{2}$, $\xi _{2}=c_{1}x+c_{3}$, $\varphi _{1}=0$, and $\varphi _{2}=-2c_{1}w$,  where $c_{1},c_{2}$ and $c_{3}$ are arbitrary constants. Consequently, the Lie algebra of the symmetry generators corresponding to the PDE system (\ref{eq:14}) is spanned by: 
\begin{equation}
\qquad X_{1}=\partial_t,\qquad X_{2}=\partial_x,\qquad X_{3}=x\partial_x-2w\,\partial_w.
\end{equation}%
%
%
%%%%%%%%%%%%%%%%%%%%%%%%%%%%%%%%%%%%%%%%%%%%%%%%%%%%%%%%%%%%%%%%%%%%%%%%%%%%%
%%%%%%%%%%%%%%%%%%%%%%%%%%%%%%%%%%%%%%%%%%%%%%%%%%%%%%%%%%%%%%%%%%%%%%%%%%%%%%%%%
\subsubsection{Approximate invariant solutions}
Now, we obtain the approximate invariant solutions corresponding to the perturbed equation (\ref{eq:13}). For the symmetry operator $X_{3}$ the corresponding characteristic equation is given by $dx/x=dt/0=dv/0=dw/(-2w)$. So, the invariants are resulted as $t=T$, $v=f(T)$, and $w=g(T)/x^2$. By inserting these invariants into the first equation of the PDE system (\ref{eq:14}), the reduced equation is $f^{\prime 2}(T)(1-f(T))=0$. Therefore, we have $v=f(T)=1/{\rm W}\big(-e^{-t-1}/c_{1}\big)$, where the function ${\rm W}(z)$ is defined implicitly by this equation $z={\rm W}(z)\,e^{{\rm W}(z)}$. After substituting this resulted solution into the second equation of the PDE system (\ref{eq:14}), we obtain $g^{\prime 2}=0$. The solution of the above equation is: 
\begin{equation}
g(T)=\frac{c_{2}\exp\big(-2{\rm W}\big(-e^{-t-1}/c_{1}\big)\big){\rm W}\big(-e^{-t-1}/c_{1}\big)}{{\rm W}\big(-e^{-t-1}/c_{1}\big)+1}.
\end{equation}
Finally, the following approximate invariant solution for the equation (\ref{eq:13}) is deduced: 
\begin{equation}
u(x,t)=f(T)+\varepsilon \frac{c_{2}\exp\big(-2{\rm W}\big(-e^{-t-1}/c_{1}\big)\big) {\rm W}\big(-e^{-t-1}/c_{1}\big) }{x^{2}\,\Big({\rm W}\big(-e^{-t-1}/c_{1}\big) +1\Big)}.
\end{equation}
%%%%%%%%%%%%%%%%%%%%%%%%%%%%%%%%%%%%%%%%%%%%%%%%%%%%%%%%%%%%%%%%%%%%%%%%%%
%%%%%%%%%%%%%%%%%%%%%%%%%%%%%%%%%%%%%%%%%%%%%%%%%%%%%%%%%%%%%%%%%%%%%%%%%%%%%%%%%%%%
\subsection{Perturbed NSW equation}
Similar to the previous sections, we will analyze the symmetries of the
perturbed NSW equation: 
\begin{equation}
u_{t}-\varepsilon u_{xx}=u(1-u^{2}).  \label{eq:15}
\end{equation}%
By applying the same calculations on this equation, the approximate symmetries are resulted as $X_{1}=\partial_t$, $X_{2}=\partial_x$, and $X_{3}=x\partial_x-2w\partial_w$. The Lie invariants corresponding to the symmetry operator $X_{3}$ are as  
$t=T,$ $v=f(T)$, and $w=g(T)/x^2$. Consequently, the following approximate invariant solution is deduced: 
\begin{equation}
u(x,t)=\frac{\pm 1}{\sqrt{1+c_{1}e^{-2t}}}+\varepsilon \frac{c_{2}e^{-2t}}{\left( 1+c_{1}e^{-2t}\right) ^{3/2}}.
\end{equation}
%
%
%%%%%%%%%%%%%%%%%%%%%%%%%%%%%%%%%%%%%%%%%%%%%%%%%%%%%%%%%%%%%%%%%%%%%%%%%%%%%%%%%%%%%%%
%%%%%%%%%%%%%%%%%%%%%%%%%%%%%%%%%%%%%%%%%%%%%%%%%%%%%%%%%%%%%%%%%%%%%%%%%%%%%%%%%%%%%%%%%%%%%%%%
\section{Optimal system of the KPP equation}
In this section, an optimal system of subalgebras corresponding to the resulted approximate symmetries of the KPP equation is constructed. As it was shown in the previous sections, the Lie algebra of the approximate symmetries corresponding to the Fisher's equation, Zeldovich equation and Newell-Whiehead-Segel (NSW) equation is three dimensional and spanned by the following generators: 
\begin{equation}  \label{eq:16}
X_{1}=\partial_t, \qquad X_{2}=\partial_x, \qquad X_{3}=x\partial_x-2w\partial_w.
\end{equation}
The commutation relations corresponding to these vector fields are given in Table 1. 
\begin{center}
{\bf Table1:} The commutator table of the \\[0pt]
approximate symmetries of the KPP equation. \\[2mm]
\begin{tabular}{ccccc}
$\left[ X_{i}, X_{j}\right] $ & $X_{1}$ & $X_{2}$ & $X_{3}$ &  \\ \hline
$X_{1}$ & $0$ & $0$ & $0$ &  \\ 
$X_{2}$ & $0$ & $0$ & $X_{2}$ &  \\ 
$X_{3}$ & $0$ & -$X_{2}$ & $0$ & 
\end{tabular}
\end{center}
It is worth noting that each $s-$parameter subgroup corresponds to one of the group invariant solutions. Since any linear combination of the
infinitesimal generators is also an infinitesimal generator, there are always infinitely many distinct symmetry subgroups for a differential
equation. But it's not practical to find the list of all group invariant solutions of a system; Consequently, we need an effective and systematic
means of classifying these solutions, leading to an \textquotedblleft optimal system" of group-invariant solutions from which every other such
solutions can be resulted. Let $G$ be a Lie group and ${\rm {\bf g}}$ denotes its Lie algebra. An optimal system of $s-$parameter subgroups is
indeed a list of conjugacy inequivalent s-parameter subgroups with the property that any other subgroup is conjugate to precisely one subgroup in
the list. Similarly, a list of s-parameter subalgebras forms an optimal system if every $s-$parameter subalgebra of ${\rm {\bf g}}$ is equivalent to
a unique member of the list under some element of the adjoint representation: $\widetilde{h}={\rm Ad}_{g}(h),$ with $g\in G$.

According to the proposition (3.7) of \cite{[5]}, the problem of finding an optimal system of subgroups is equivalent to that of obtaining an optimal system of subalgebras. For one-dimensional subalgebras, this classification problem is essentially the same as the problem of classifying the orbits of the adjoint representation. Since each one-dimensional subalgebra is determined by a nonzero vector in ${\rm {\bf g}}$, this problem is attacked by the naive approach of taking a general element $X$ in ${\rm {\bf g}}$ and subjecting it to various adjoint transformations so as to simplify it as much as possible. Thus we will deal with the construction of an optimal system of subalgebras of ${\rm {\bf g}}$. The adjoint action is given by the Lie series: ${\rm Ad}(\exp (\varepsilon X_{i},X_{j})=X_{j}-\varepsilon \lbrack X_{i},X_{j}]+\varepsilon^2/2)[X_{i},[X_{i},X_{j}]]-\cdots$, where $[X_{i},X_{j}]$ denotes the Lie bracket, $\varepsilon $ is a parameter and $i,j=1,2,3$ (\cite{[5]}).

The adjoint representation ${\rm Ad}$ corresponding to the resulted approximate symmetries is presented in table 2 with the $(i, j)$-th entry indicating ${\rm Ad}(\exp(\epsilon x_{i})x_{j})$.
\begin{center}
{\bf Table2:} Adjoint representation of the \\[0pt]
approximate symmetries of the KPP equation. \\[2mm]
\begin{tabular}{ccccc}
${\rm Ad}$ & $X_{1}$ & $X_{2}$ & $X_{3}$ &  \\ \hline
$X_{1}$ & $X_{1}$ & $X_{2}$ & $X_{3}$ &  \\ 
$X_{2}$ & $X_{1}$ & $X_{2}$ & $X_{3}-\varepsilon X_{2}$ &  \\ 
$X_{3}$ & $X_{1}$ & $e^{\varepsilon }X_{2}$ & $X_{3}$ & 
\end{tabular}
\end{center}
Therefore, we can state the following theorem:
\paragraph{Theorem 1:}
An optimal system of one dimensional subalgebras corresponding to the Lie
algebra of approximate symmetries of the KPP equation is generated by: (i) $X_{1}$, (ii) $\alpha X_{1}+X_{2}$, (iii) $\beta X_{1}+X_{3}$, 
where $\alpha, \beta \in {\rm {\bf R}}$ are arbitrary constants.
\paragraph{Proof:}
Let $F_{i}^{s}:{\rm {\bf g}}\longrightarrow {\rm {\bf g}}$ be a linear map
defined by $X\longrightarrow {\rm Ad}(\exp (s_{i}X_{i})X)$ for $i=1,\cdots
,3 $. The matrices $M_{i}^{s}$ of $F_{i}^{S}$ with respect to the basis $%
\left\{ X_{1},X_{2},X_{3}\right\} $ are given by: 
\begin{equation}
M_{1}^{s}=\left( 
\begin{array}{ccc}
1 & 0 & 0 \\ 
0 & 1 & 0 \\ 
0 & 0 & 1%
\end{array}%
\right) ,\quad M_{2}^{s}=\left( 
\begin{array}{ccc}
1 & 0 & 0 \\ 
0 & 1 & 0 \\ 
0 & -s_{1} & 1%
\end{array}%
\right) ,\quad M_{3}^{s}=\left( 
\begin{array}{ccc}
1 & 0 & 0 \\ 
0 & e^{s_{2}} & 0 \\ 
0 & 0 & 1%
\end{array}%
\right) .
\end{equation}%
Let $X=$ $\sum\limits_{i=1}^{3}a_{i}X_{i}$, then  $F_{3}^{s}\circ F_{2}^{s}\circ F_{1}^{s}:X\longmapsto a_{1}X_{1}+a_{2}e^{s_{2}}X_{2}+(a_{3}-s_{1}a_{2})X_{3}$. In the following, by alternative action of these matrices on a vector field $%
X$, the coefficients $a_{i}$ of $X$ will be simplified.

If $a_{2}\neq 0, $ then we can make the coefficients of $X_{3}$ vanish by $F_{1}^{s}$; By setting $s_{1}=a_3/a_2$. Scaling $X$ if necessary, we can assume that $a_{2}=1$. So, $X$ \ is reduced to the case (ii). If $a_{2}=0$ and $a_{3}\neq 0$, by scaling we insert $a_{3}=1$. So $X$ is reduced to the case (iii). Finally, if $a_{2}=a_{3}=0, $ then $X$  is reduced to the case (i). There is not any more possible cases for investigating and the proof is complete.

In order to obtain the group transformations which are generated by the
resulted infinitesimal symmetry generators (\ref{eq:16}), we need to solve
the following system of first order ordinary differential equations $%
(x_{1}=x, x_{2}=t, u_{1}=v, u_{2}=w)$. 
\begin{eqnarray}
d\widetilde{x}_{j}(s)/ds &=&\xi _{i}^{j}(\widetilde{x}(s), \widetilde{t}(s), \widetilde{v}(s), \widetilde{w}(s)), \qquad \widetilde{x}_{j}(0)=x_{j}, \qquad i=1, 2, 3. \qquad  \nonumber \\
d\widetilde{u}_{j}(s)/ds &=&\varphi _{i}^{j}(\widetilde{x}(s), \widetilde{t}(s), \widetilde{v}(s), \widetilde{w}(s)), \qquad \widetilde{u}_{j}(0)=u_{j}, \qquad j=1, 2.
\end{eqnarray}
Hence, by exponentiating the resulted infinitesimal approximate symmetries
of the KPP equation, the one-parameter groups $G_{i}(s)$ generated by $X_{i}$
for $i=1, 2, 3 $ are determined as follows: 
\begin{eqnarray}
G_{1} &:&(t, x, v, w)\longmapsto (t+s, x, v, w),  \nonumber \\
G_{2} &:&(t, x, v, w)\longmapsto (t, x+s, v, w), \\
G_{3} &:&(t, x, v, w)\longmapsto (t, e^{s}x, v, e^{-2s}w).  \nonumber
\end{eqnarray}
Consequently, we can state the following theorem:
\paragraph{Theorem 2:}
If $u=f(t, x)+\varepsilon g(t, x)$ is a solution of the KPP equation, so are
the following functions: 
\begin{eqnarray}
G_{1}(s)\cdot u(t, x) &=&f(t-s, x)+\varepsilon g(t-s, x),  \nonumber \\
G_{2}(s)\cdot u(t, x) &=&f(t, x-s)+\varepsilon g(t, x-s), \\
G_{3}(s)\cdot u(t, x) &=&f(t, e^{-s}x)+\varepsilon e^{-2s}g(t, e^{-s}x). 
\nonumber
\end{eqnarray}
%%%%%%%%%%%%%%%%%%%%%%%%%%%%%%%%%%%%%%%%%%%%%%%%%%%%%%%%%%%%%%%%%%%%%%%%%%%%%%%%%%%%
%%%%%%%%%%%%%%%%%%%%%%%%%%%%%%%%%%%%%%%%%%%%%%%%%%%%%%%%%%%%%%%%%%%%%%%%%%%%%%%%%%%%
\section*{Conclusion}
The investigation of the exact solutions of nonlinear PDEs plays an
essential role in the analysis of nonlinear phenomena. Lie symmetry method
greatly simplifies many nonlinear problems. Exact solutions are nevertheless
hard to investigate in general. Furthermore, many PDEs in application depend
on a small parameter, hence it is of great significance and interest to
obtain approximate solutions. Perturbation analysis method was thus
developed and it has a significant role in nonlinear science, particularly
in obtaining approximate analytical solutions for perturbed PDEs. This
procedure is mainly based on the expansion of the dependent variables
asymptotically in terms of a small parameter. The combination of Lie group
theory and perturbation theory yields two distinct approximate symmetry
methods. The first method due to Baikov et al. generalizes symmetry group
generators to perturbation forms \cite{[baikov],[1]}. The second method
proposed by Fushchich and Shtelen \cite{[3]} is based on the perturbation of
dependent variables in perturbation series and the approximate symmetry of
the original equation is decomposed into an exact symmetry of the system
resulted from the perturbation. Taking into account the comparison in \cite%
{[compare]} the second method is superior to the first one.\newline
As it is well known, the solutions of nonlinear reaction-diffusion equations
represent a wide class of behaviors, including the formation of wave-like
phenomena and traveling waves as well as other self-organized patterns. In
this paper we have comprehensively analyzed the approximate symmetries of a
significant class of nonlinear reaction-diffusion equations called
Kolmogorov-Petrovsky-Piskounov (KPP) equation. For this purpose, we have
concentrated on four particular and important forms of this equation
including: Fick's second law, Fisher's equation, Zeldovich equation and
Newell-Whitehead-Segel (NWS) equation. It is worth mentioning that in order
to calculate the approximate symmetries corresponding to these equations, we
have applied the second approximate symmetry method which was proposed by
Fushchich and Shtelen. Meanwhile, we have constructed an optimal system of
subalgebras. Also, we have obtained the symmetry transformations and some
invariant solutions corresponding to the resulted symmetries.
\section*{Acknowledgements}
The authors are wish to thank the Mis. Fatemeh Ahangari for careful reading
and useful suggestions. 
%%%%%%%%%%%%%%%%%%%%%%%%%%%%%%%%%%%%%%%%%%%%%%%%%%%%%%%%%%%%%%%%%%%
%%%%%%%%%%%%%%%%%%%%%%%%%%%%%%%%%%%%%%%%%%%%%%%%%%%%%%%%%%%%%%%%%%%

\end{document}